\documentclass[reprint,
superscriptaddress,
 amsmath,amssymb,
 aps,
]{revtex4-1}

\usepackage{graphicx}
\usepackage{dcolumn}
\usepackage{bm}
\usepackage{enumerate}
\usepackage{lipsum}
\usepackage{epstopdf}
\usepackage{float}
\usepackage[caption = false]{subfig}

\begin{document}

\preprint{APS/123-QED}

\title{The optical manipulation of matter-wave vortices: An analogue of circular dichroism}

\author{Pradip Kumar Mondal}
\affiliation{Department of Physics, Indian Institute of Technology Kharagpur, Kharagpur-721302, India.}
\author{Bimalendu Deb}%
\affiliation{Department of Materials Science, Indian Association for the Cultivation of Science, Jadavpur, Kolkata 700032, India.}
\author{Sonjoy Majumder}
\email{sonjoym@phy.iitkgp.ernet.in}
\affiliation{Department of Physics, Indian Institute of Technology Kharagpur, Kharagpur-721302, India.}





\begin{abstract}

The transfer of orbital angular momentum from an optical vortex to an atomic Bose-Einstein condensate changes the vorticity of the condensate. The spatial mismatch between initial and final center-of-mass wavefunctions of the condensate influences significantly the two-photon optical dipole transition between corresponding states. We show that the transition rate depends on the handedness of the optical orbital angular momentum leading to optical manipulation of matter-wave vortices and circular dichroism-like effect. Based on this effect, we propose a method to detect the presence and sign of matter-wave vortex of atomic superfluids. Only a portion of the condensate is used in the proposed detection method leaving the rest in its initial state.
\end{abstract}

\maketitle

\section{INTRODUCTION}
The interaction of chiral molecules with light is sensitive to the circular polarization or helicity of the photons. One manifestation of this optical activity is circular dichroism (CD), i.e., light absorption in such material is sensitive to the handedness of the circular polarization. Such interaction is enantiometrically specific and depends on the structure of the chiral matter \cite{barron_04}. Polarization is associated with the spin angular momentum (SAM)of light. Therefore, CD effect is usually associated with spin of light. The work of Allen \textit{et al.} \cite{allen_92} has shown that in addition to SAM, Laguerre-Gaussian (LG) beams carry well defined orbital angular momentum (OAM) associated with its spatial mode. This has triggered new research on interaction of matter with laser beams having certain spatial profiles, like, LG or Bessel profile \cite{allen_03,andrews_13}. From extensive theoretical \cite{andrews_04, andrews_12, coles_12} and experimental works \cite{araoka_05,loffler_11} it has been believed that the OAM of LG beam does not play a role in CD. On the contrary, here we prove that the matter-wave vortex of a Bose-Einstein condensate (BEC) interacting with an LG beam can lead to CD-like effects, i.e., the light absorption becomes sensitive to the handedness of the field OAM.

Numerous theoretical and experimental studies related to creation of matter-wave vortex states, persistent current in BEC and coherent control of the OAM of atoms using interaction of ultra-cold atoms with optical vortex \cite{nandi_04,andersen_06, simula_08, song_09, wright_08, wright_09, jaouadi_10, gullo_10, tasgin_11, lembessis_11, brachmann_11, kanamoto_11, ramanathan_11, robb_12, okulov_12, beattie_13} have been reported over past two decades. Quantized vortex states play an essential role in macroscopic quantum phenomena like superfluidity and superconductivity. There have been studies to detect sign of a vortex state using different techniques, like, interference between vortex states \cite{bolda_98, chevy_01}, exciting the quadrupole mode of a BEC using an auxiliary laser beam as stirrer \cite{chevy_00}, vortex core precession \cite{anderson_00,freilich_10}, method of twisted densities \cite{donadello_14}, and vortex gyroscope imaging (VGI) \cite{powis_14}. The primary difficulty of in-situ observation of vortices in trapped condensate is that the radius of a vortex core is on the order of the condensate healing length, which is generally several times smaller than the wavelength of the laser used for imaging. Methods applying free expansion after extinguishing the trap or interferometric steps disrupt the time evolution irreversibly. In most of the experiments, reproducing precisely the initial condition of vortex generation is very challenging. In some methods, like, rapid quench across the BEC transition during evaporation via Kibble-Zurek mechanism (KZM) \cite{kibble_76,zurek_85}, the vortex generation is even stochastic \cite{weiler_08}. So, any detection method using a fraction of the condensate \cite{freilich_10} or in-situ nondestructive phase-contrast imaging \cite{anderson_00} is always preferable to any destructive method. 

Here we show how the CD-like effect that arises in interaction of a BEC with an LG beam can be useful to detect a single vortex of BEC and its handedness keeping a portion of  initial BEC undisturbed. In this method, two-photon stimulated Raman transitions are used only in a portion of BEC that is spatially separated out from the rest of the condensate which remains trapped in its initial state. One can detect the presence and sign of the undisturbed vortex by doing measurements on the separated part. In recent experiments of \cite{andersen_06,wright_08,wright_09}, two-photon transitions are used to transfer the OAM to the atomic BEC employing an LG and a Gaussian beam.  

\section{THEORY}
The mechanism of angular momentum transfer from an LG beam to ultracold atoms is studied in paraxial limit in our previous work \cite{mondal_14}. Here we recall some of the essential features of this OAM transfer mechanism. We consider an LG beam without any off-axis node propagating along $ z $ axis of the laboratory frame interacting with $ ^{23} $Na BEC whose de Broglie wavelength is large enough to feel the intensity variation of LG beam but smaller than the waist of the beam. Under this condition, the dipole interaction Hamiltonian can be written as
\begin{widetext}
\begin{align}
H_{\text{I}}=\sqrt{\dfrac{4\pi }{3\vert l\vert !}} e\left(\dfrac{1}{w_0}\right)^{\vert l\vert} r\sum_{\sigma =0,\pm1}\epsilon_{\sigma}Y_1^{\sigma}({\bf\hat{\textbf{r}}})R_{\text{c.m.}\perp}^{\vert l \vert } e^{il\Phi_{\text{c.m.}}}e^{ikZ_{\text{c.m.}}} + \text{H.c.}
\end{align}
\end{widetext}
where $ l $ and $ w_0 $ are the winding number and the waist of the LG beam. $ \textbf{r} $ is coordinate of the valence electron with respect to center-of-mass (c.m.) of the atom and $ \textbf{R} $ is the coordinate of the c.m. in laboratory frame of reference. Rabi frequency is given by $ \Omega_{\text{I}} = \frac{1}{\hbar }\langle \Upsilon_f\vert H_{\text{I}}\vert \Upsilon_i\rangle  $, where $ \Upsilon $ denotes an unperturbed atomic state, $ \Upsilon (\textbf{R}_{\text{c.m.}},\textbf{r}) = \Psi_{\text{c.m.}}(\textbf{R}_{\text{c.m.}})\psi_e (\textbf{r})   $ with $ \Psi_{\text{c.m.}}(\textbf{R}_{\text{c.m.}}) $ being the c.m. wavefunction that depends on the external trapping potential and $ \psi_e (\textbf{r}) $ is the internal electronic wavefunction. $ \Psi_{\text{c.m.}}(\textbf{R}_{\text{c.m.}}) $ is calculated by solving the Gross-Pitaevskii equation \cite{mondal_14} and $ \psi_e (\textbf{r}) $ may be taken as a correlated orbital obtained from many-body calculation \cite{mondal_13}. Detail calculations are given in our previous work \cite{mondal_14}. There we have proved that electric dipole interaction of LG beam with BEC allows the transfer of the total field OAM to the c.m. motion of BEC and thus, changes the vorticity of the matter-wave state. However, the SAM corresponding to polarization is transferred to the internal electronic motion and controls the selection rule.

We consider sodium BEC initially prepared in state $ \vert \kappa; F,m_F\rangle $ where $ \kappa $ is vorticity of c.m. state and $ F $, $ m_F $ correspond to hyperfine spin of ground-state Na atoms. Figure 1 (a) shows a stimulated Raman scheme using two sets of counter propagating LG and G pulses. An atom of mass $ M $, initially at rest, absorbs an LG photon and stimulatedly emits a G photon, acquiring $ 2\hbar k $ of linear momentum (LM) ($ k=2\pi / \lambda $ with $ \lambda $ the photon wavelength) in the direction of propagation the LG photon. Here we reason out how the two-photon Rabi frequency will depend on the handedness of the OAM of the LG beam. Two-photon detuning $ \Delta $ is much larger than the spontaneous emission linewidth of the excited state. After adiabatically eliminating the intermediate excited state, the effective Hamiltonian of the system of two-coupled vortices can be written as $ H_{\pm}=H_{\pm}^0+H_{\pm}' $ where
\begin{align}
 H_{\pm}^0 =\hbar  \kappa \omega_i \vert \kappa; \eta \rangle \langle \kappa; \eta \vert +  
 \left (\hbar \omega_f + \epsilon_q \right) \vert \kappa \pm l; \eta' \rangle \langle \kappa \pm l; \eta' \vert 
\end{align}
and
\begin{align}
 H_{\pm}'= \left [ \Omega_{\pm} e^{-i \delta_{\pm} t} \vert \kappa \pm l; \eta' \rangle \langle \kappa; \eta\vert + {\rm C. c.} \right ].
\end{align}
where $ \eta = \left\lbrace F,m_F\right\rbrace $ and $ \eta' = \left\lbrace F',m_F'\right\rbrace $ are spin states of un-scattered and Raman-scattered atoms, respectively. The vorticity of the scattered atoms is $ \kappa \pm l $. $ \omega_i $ ($ \omega_f $) is frequency corresponding to the total energy (c.m. + internal) of initial (final) state. $\Omega_{\pm}$ is the two-photon Rabi frequency for LG$_0^{\pm l}$/G pulse and $\delta_{\pm} = \omega_{{\rm LG}} - \omega_{{\rm G}}$ is the detuning between the two beams. The states $\vert \kappa \pm l; F',m_F' \rangle$ receive the momentum transfer ${\mathbf q} = {\mathbf k}_{{\rm LG}} -  {\mathbf k}_{{\rm G}}$ due to two-photon stimulated light scattering. As a result, the atoms in  $\vert \kappa \pm l; F',m_F'\rangle$ gain kinetic energy $\epsilon_q = (\hbar q)^2/2M$ where $ M $ is mass of an atom. The important point here is that the two-photon Rabi frequency $\Omega_{+} = \Omega_G \Omega_{\text{I}}^{+}/\Delta$ for LG$_0^{+l}$/G pulse is different from  $\Omega_{-}$ for LG$_0^{-l}$/G pulse. The radial portions of c.m. wavefunctions of BEC ($\Psi_{\text{c.m.}}(\textbf{R}_{\text{c.m.}})$) corresponding to vorticities $\kappa + l  $ and $ \kappa -l $ are different (see Figure 2 of \cite{mondal_14}). This makes the Rabi frequencies $ \Omega_+ $ and $ \Omega_- $ corresponding to the two transitions different. Thus, in principle, a BEC vortex state is expected to show CD-like behavior in interaction with LG beams having OAM of opposite handedness. However, if initially the BEC was in a non-vortex state ($ \kappa = 0 $) then after interaction with LG$ _0^{+l} $/G and LG$ _0^{-l} $/G pulses for an appropriate pulse duration the final states of BEC will have vorticities $ +l $ and $ -l $, respectively. The radial portions of c.m. wavefunctions of these two states are identical. This makes the c.m. matrix elements and Rabi frequencies corresponding to these transitions identical. Hence, non-vortex state in BEC does not show CD-like behavior in interaction with LG beam. 

The Hamiltonian $ H_{\pm } $ given by Eqs. (2) and (3) describes coherent dynamics of spin-vortex coupled states. Let the solutions of this Hamiltonian be 
\begin{align}
\vert \psi_{\pm}(t) \rangle = a_{\pm} (t) \vert \kappa;F,m_F \rangle + b_{\pm}(t) \vert \kappa \pm l; F',m_F' \rangle.
\end{align}
Under phase matching condition $\epsilon_q = \hbar \delta_{\pm}$ and assuming $ (\omega_f-\omega_i) < \! < \Omega_{\pm}$ we have 
\begin{align}
|a_{\pm}(t)|^2 = \cos^2 \left (\Omega_{\pm} t \right )
\end{align}
and 
\begin{align}
|b_{\pm}(t)|^2 = \sin^2 \left (\Omega_{\pm} t  \right ) .
\end{align}
This shows that for pulse duration $t_p$ given by $ \Omega_{\pm} t_p = \pi/2$, the vortex state  $\vert\kappa \rangle$ will be coherently transferred into the vortex state  $\vert \kappa \pm l;F',m_F' \rangle$.  For  $ \Omega_{\pm} t_p = \pi/4$ there will be equal superposition of the two vortex states. Thus our scheme opens up new prospect for optical manipulation of matter-wave vortices with an effect that is analogous to CD. It is important to note that the time evolution of the two states $ \psi_+ $ and $ \psi_- $ are different only due to CD-like effect that arises when $ \kappa \ne 0 $ \textit{i.e.}, if the BEC is initially in a vortex state. On the other hand, the temporal evolution of the two states will be the same for $ \kappa =0 $ due to the absence of CD-like effect. Hence, this effect will play an important role in quantum evolution of spin-vortex coupled BEC in a spinor condensate. The state given by Eq. (4) is an entangled state of spin and matter-wave vortices.

\section{PROPOSED APPLICATION}
\begin{figure}
\subfloat[]{\includegraphics[trim = 1cm 4cm 1cm 2cm, scale=.4]{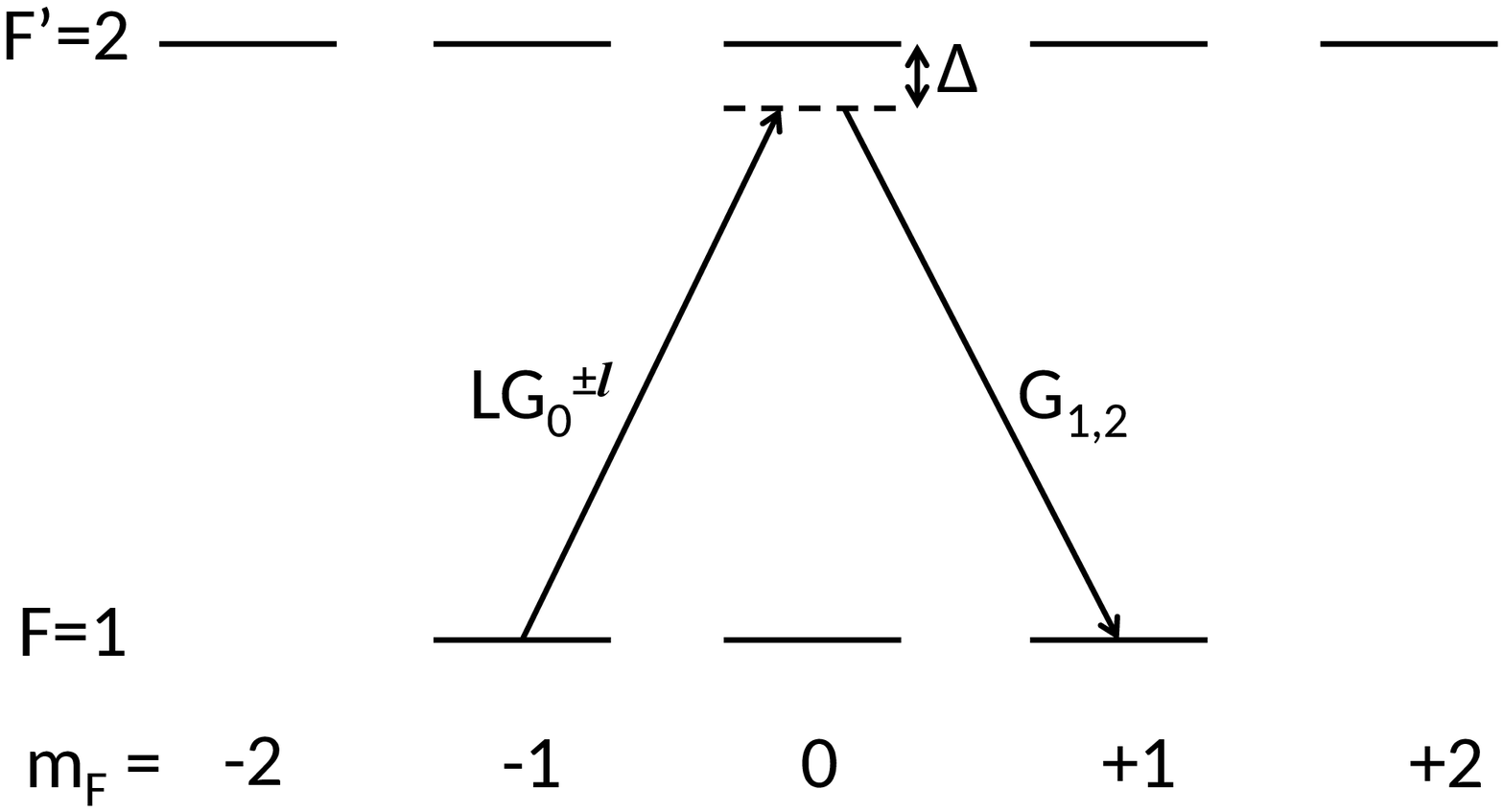}}\\
\subfloat[]{\includegraphics[trim = 1cm 4cm 1cm 0cm, scale=.35]{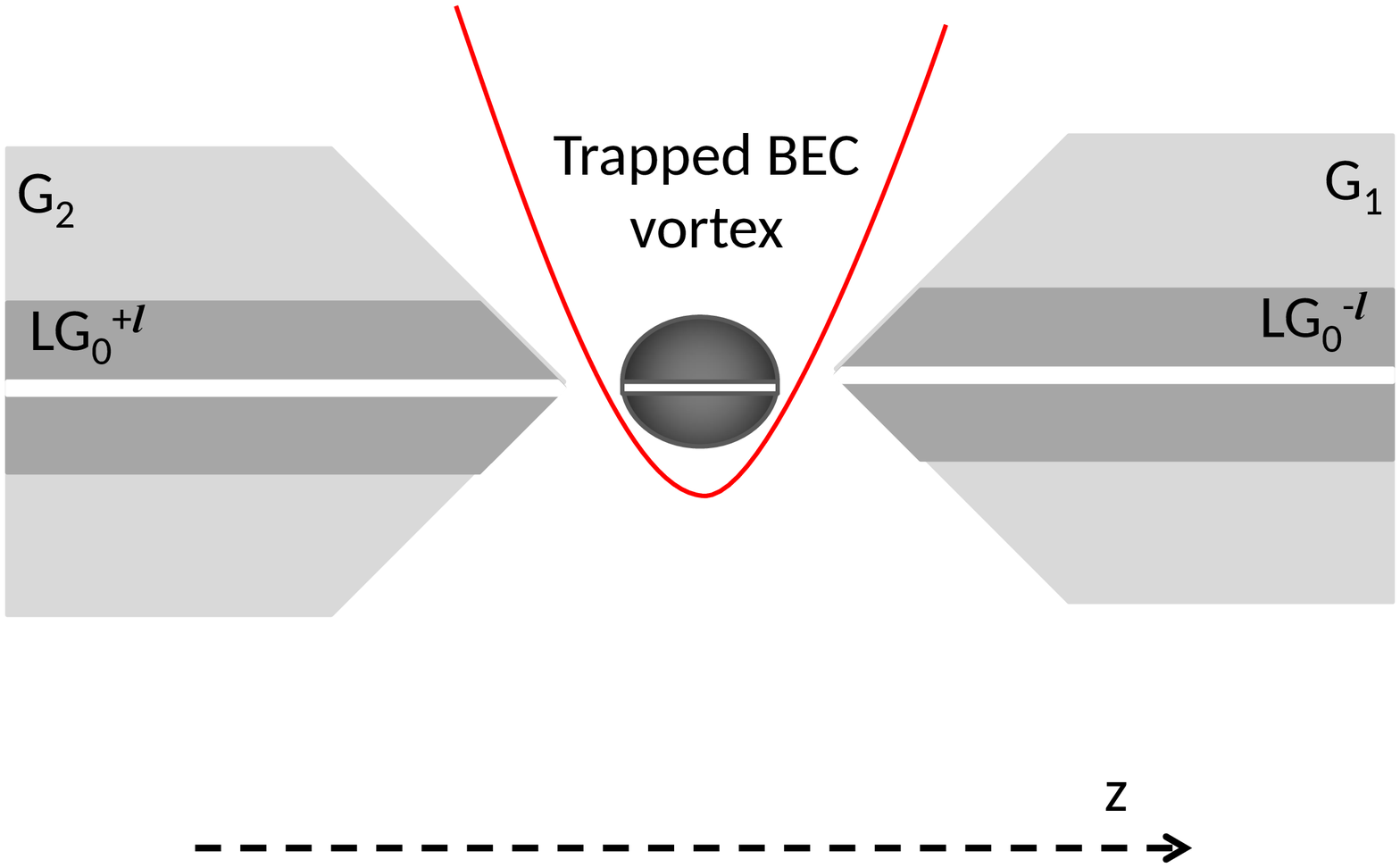}}\\
\subfloat[]{\includegraphics[trim = 1cm 4cm 1cm 0cm,scale=.35]{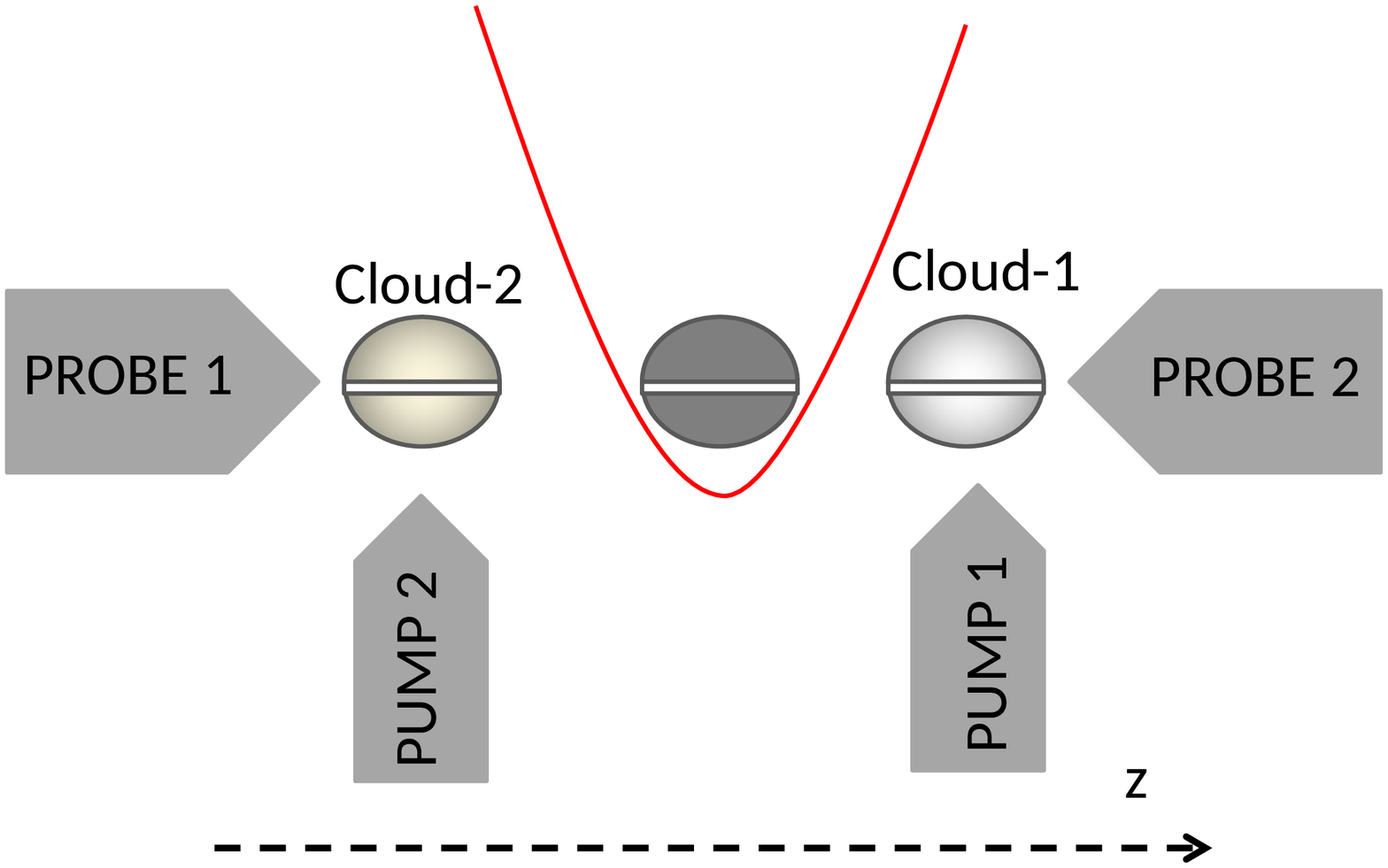}}
\caption{Proposed experimental scheme illustrating the process to detect handedness of matter-wave vortex of an atomic Bose-Einstein condensate using the CD-like effect. (a) Energy level scheme of the two-photon transitions. The atomic states shown correspond to the $ ^{23} $Na hyperfine structure. Atoms in the BEC are initially in  $\vert 3S_{1/2},F=1,m_F=-1\rangle $ electronic state. Afte 2-photon transitions the final electronic states become $ \vert 3S_{1/2},F=1,m_F=1\rangle $. $ \Delta $ denotes two-photon detuning (b) LG$ _0^{l} $/G$ _1 $ and LG$ _0^{-l} $/G$ _2 $ pulses are simultaneously applied to the BEC. (c) The atoms that have undergone the type-1 Raman transitions have come out as cloud-1 and those which have undergone the type-2 Raman transitions have separated as cloud-2 after a few millisecond TOF. Two pump beams along with two probe beams enable independent imaging of the two clouds (for details, see the text).}
\end{figure}

 Now, we discuss how this CD-like effect can be useful to detect vortex state in BEC and determine the handedness of vorticity using two LG beams having OAM of opposite handedness. We consider sodium BEC is initially prepared in electronic state $ \vert 3S_{1/2},F=1,m_F=-1\rangle $ \cite{andersen_06}. Our proposed experimental scheme is shown in figure 1. LG$ _0^{l} $ and G$ _2 $ pulses propagate along $ +Z $ axis and LG$ _0^{-l} $ and G$ _1 $ pulses propagate along $ -Z $. We apply LG$ _0^{l} $/G$ _1 $ and LG$ _0^{-l} $/G$ _2 $ pulses simultaneously. The two-photon Raman transition by LG$ _0^{l} $/G$ _1 $ pulse is called type-1 transition and the other two-photon Raman transition by LG$ _0^{-l} $/G$ _2 $ pulse is called type-2 transition in this paper. The atoms taking part in the Raman transitions for a pulse duration of the order $ \Omega_{\pm}^{-1} $ will be in final spin state $ \vert 3S_{1/2},F=1,m_F=+1\rangle $ which is high field seeking state. Therefore, they will no longer be trapped. In addition, these atoms will gain $ 2\hbar k $ LM from the Raman process and propagate ballistically. After a few millisecond time of flight (TOF), these atoms will be spatially separated from the atoms which are still at rest and trapped. Let us call the atom cloud which has gone through type-1 transition as cloud-1 and the other one as cloud-2. Then one can use focused pump beam spatially localized along $ z $ and selectively image atom clouds in different LM states and different spatial positions. In accordance with the CD-like effect described in the previous paragraph, the number of atoms in cloud-1 will be different from that of cloud-2 depending on the handedness of the initial BEC vortex state.

\section{NUMERICAL RESULTS}
We now proceed to numerically evaluate how the number of atoms in cloud-1 will differ from that of cloud-2 depending on the sign of vorticity of the initial BEC. As mentioned in the previous paragraph, we consider sodium BEC containing $ 10^4 $ number of atoms initially prepared in electronic state  $ \vert 3S_{1/2},F=1,m_F=-1\rangle $ \cite{andersen_06}. The atoms are trapped in an anisotropic trap. The asymmetry parameter $\omega_Z / \omega_{\perp} = 2 $ with axial frequency $ \omega_Z / 2\pi = 40 $ Hz. The corresponding characteristic length is $ a_{\perp}= 4.673$ $\mu $m. $ s $- wave scattering length is $ a=2.75 $ nm \cite{simula_08, inouye_98}. For phase matching, we set the frequency difference between the two counter propagating LG and G beams as $ \delta \nu = 4E_r/h $, where $ E_r = (\hbar k)^2/2M $ is the recoil energy \cite{andersen_06} (see figure 2).
\begin{figure}
\includegraphics[trim = 2cm 3cm 1cm 0cm, scale=.45]{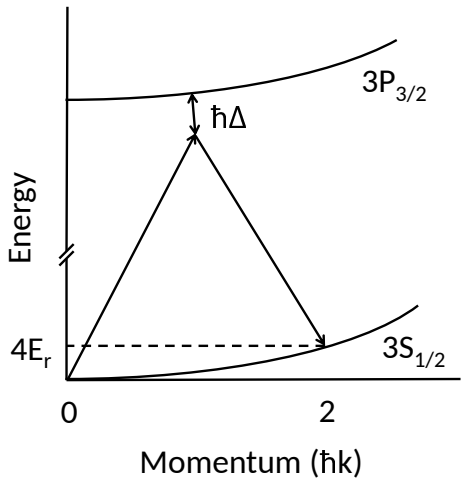}
\caption{Diagram illustrating energy and LM conservation of the 2-photon Raman process for LG/G pulse.}
\end{figure}
In addition to LM, the atoms pick up the OAM difference between the two photons and this causes change in rotational motion of the atoms. Corresponding rotational kinetic energy difference between final and initial vortex states is small compared to $ \delta \nu $ \cite{andersen_06} and gives rise to rotational Doppler shift \cite{barreiro_06}. The laser beams are detuned from the \textit{D}$_2$ line ($ \lambda = 589.0 $ nm) by $ \Delta = -1.5 $ GHz ($ \approx 150 $ linewidths, enough to prevent any significant spontaneous photon scattering). The frequency of G$ _{1,2} $ are further decreased from LG$ ^{\pm l} $ beams by $ \delta \nu $ (may be using acousto-optic modulators).  The intensity of the laser beams are set to be $ I = 10^2  $ Wcm$ ^{-2} $. The waists of the LG beams are set to be $ w_0 = 10^{-4} $m. The radius of the G$ _{1,2} $ beams is generally a order of magnitude larger than that of the BEC and hence, during interaction with the condensate we have considered the G$ _{1,2} $ beams as plane waves. After type-1 (type-2) two-photon transition, the vorticity of final c.m. state is going to be changed by $ +l $ ($ -l $).
\begin{table}
\caption{This illustrates cicular dichroism-like effect in BEC vortex states undergoing type-1 and type-2 two-photon Raman transitions. $ \kappa_i $ and $ \kappa_f $ are vorticity of initial and final states of the condensate, respectively. $ l $ denotes the winding number of the LG beam associated with the transition. $ \Omega_+ $ and $ \Omega_- $ are dipole Rabi frequencies (s$ ^{-1} $) corresponding to type-1 and type-2 transitions, respectively. }
\centering
\begin{tabular}{ccccccccccccc}
\hline \hline
$ \kappa_i $&&$ l $&&$ \kappa_f $&&$ \Omega_+ $&&$ l $&&$ \kappa_f $&&$ \Omega_- $\\
\hline
1&&+1&&2&&7.49$ \times 10^{8} $ &&-1&&0&& 6.53$ \times 10^{8} $\\
 &&+2&&3&&2.42$ \times 10^{6} $ &&-2&&-1&&2.03$ \times 10^{6} $ \\
-1&&+1&&0&& 6.53$ \times 10^{8} $&&-1&&-2&&7.49$ \times 10^{8} $\\
  &&+2&&+1&&2.03$ \times 10^{6} $ &&-2&&-3&&2.42$ \times 10^{6} $\\

\hline
\end{tabular}
\end{table}
The two-photon Rabi frequencies corresponding to type-1 and type-2 transitions are defined as $ \Omega_+ $ and $ \Omega_- $, respectively, shown in Table I. Our calculation shows that if $ \kappa_i > 0 $ then $ \Omega_+ > \Omega_- $ and if $ \kappa_i < 0 $ then $ \Omega_+<\Omega_- $. 

Andersen \textit{et. al.} \cite{andersen_06} used similar two-photon Raman transition technique to generate and image BEC vortex. They worked with BEC of $ 10^6 $ Na atoms and achieved maximum efficiency of the two-photon transition as high as $ 53\% $ for a $ 130 $ $ \mu $s LG/G pulse. From their experimental observation we can estimate roughly how the number of atoms in cloud-1 will vary from the number of atoms in cloud-2. Let us start with $ 10^4 $ number of atoms with initial vorticity $ \kappa_i = +1 $ and we consider that $ \Omega_+ $ and $ \Omega_- $ are experimentally calibrated so that they are equal for a non-vortex  initial state. We assume that the experimental parameters are set such that $ 30\% $ of atoms take part in type-2 transition. According to Table I, if $ \vert l\vert = 1$, then almost $ 40\% $ of atoms will take part in type-1 transition and that means cloud-1 will have almost $  10^3 $ number of atoms more than cloud-2. If we use LG beams with $ \vert l\vert = 2 $ then this difference in number of atoms present in cloud-1 and cloud-2 is almost $1.3\times 10^3  $ ($ 13\% $ of initial number of atoms). This difference of $ 10\% $ (for $ \vert l\vert = 1 $) number of atoms is easily detectable by absorption imaging of cloud-1 and cloud-2 \cite{ketterle, bloch} and thereby one can determine the handedness of rotation of the initial BEC. It will be same as the rest of portion of the BEC which has not taken part in the Raman processes. If the number of atoms in cloud-1 and cloud-2 are almost same then the BEC had no vortex state. 

\section{CONCLUSION}
In conclusion, we have demonstrated that CD-like effects can arise in interaction of a matter-wave vortex with LG beams of opposite OAM. This effect will have wide applications including detection of handedness of the matter-wave vortex as theoretically demonstrated in this paper. While creating a vortex state from non-rotating BEC using two-photon Raman transitions by applying square pulses of a particular duration, the transfer efficiency is limited by the spatial mismatch between the rotating state and the initial BEC \cite{andersen_06}. In this work, we have shown that it is possible to take advantage of this limitation that lies at the heart of the predicted CD-like effect. As an application of this effect we have proposed a method to detect BEC vortex states and its handedness using a portion of the condensate. The main advantages of this method are as follows. Firstly, only single imaging of cloud-1 and cloud-2 is needed. Secondly, we do not need to know the exact number of atoms present in the clouds. All we need to know is which of the clouds contains more number of atoms. Thirdly, the rest of the condensate which does not take part in the two-photon transitions will remain trapped in its initial state. While this method is suitable for steadily rotating condensate containing single vortex, the situation is different for multiply quantized vortices in a BEC. It needs further studies to extend the method for many-vortex configurations.

\section*{ACKNOWLEDGMENT}
The authors acknowledge useful discussions with Dr. Narendra Nath Dutta from Indian Institute of Technology Kharagpur, India. P.K.M. acknowledges financial support from the Council of Scientific and Industrial Research (CSIR), India.

\end{document}